\documentclass[pdflatex,sn-basic]{sn-jnl}
\usepackage{graphicx}
\usepackage{amsmath,amssymb}
\usepackage{booktabs}
\usepackage{xcolor}
\graphicspath{{./}}
\newcommand{\papertitle}{Do links to retracted papers predict future retractions? An out-of-sample test on the open biomedical citation graph}
\newcommand{\papershorttitle}{Retracted neighbours and future retractions}
\newcommand{\paperkeywords}{retraction, research integrity, citation networks, paper mills, OpenCitations, Retraction Watch}
\begin{document}

\title[\papershorttitle]{\papertitle}

\author*[1]{\fnm{Hossein} \sur{Mohasel Arjomandi}}\email{hosseinm@microsoft.com}
\equalcont{The views expressed in this article are the author's own and do not necessarily reflect those of Microsoft.}
\affil*[1]{\orgname{Microsoft}, \orgaddress{\city{Redmond}, \state{WA}, \country{USA}}}

\abstract{%
Tools that screen manuscripts for research-integrity problems flag papers that cite retracted work, and retrospective
studies report that such papers are retracted more often. This study tests the signal prospectively on open data.
Predictors were computed on the January 2024 OpenCitations graph using the retractions known at that date, frozen, and
compared with retractions in the Retraction Watch database from February 2024 to October 2026. The design was
pre-specified. Among 10.5 million biomedical papers published 2015--2023, 2.1\% cited a paper already retracted, and
4,323 were retracted later (0.041\%). Citing a retracted paper raised the later retraction rate from 0.035\% to 0.27\%
with one such reference; excluding reviews (post hoc), it rose to 2.3\% with five or more. Within journal and publication year the odds ratio was 2.99 (95\%
CI 2.73--3.27), and 3.03 when reference-list length was also held fixed (post hoc), so the signal is not explained by
retraction waves in journals or by long reference lists. As a screen it is weak: 0.9\% of the
1,000 highest-ranked papers were retracted later (2.2\% with reviews set aside, post hoc), and 84\% of later retractions cite
no retracted paper. When the time
point was moved into the late-2023 retraction wave, the journal's own retraction rate became the far better screen.
NIH-funded papers cited already-retracted work as often as other papers in the same journal and year. Links to
retracted papers are a useful flag but not a screen on their own.}


\maketitle

\section{Introduction}\label{sec:intro}

Retractions have become far more frequent. Counts computed from the Retraction Watch database (copy of 30 October 2026) show that 13,210 papers
have a first retraction notice dated 2023, more than in any earlier year, and papers retracted for paper-mill reasons rose from 246 in 2020 to
6,782 in 2023
\citep{rittman2025rw,candal2022papermills}. Because retractions are slow, often years after publication, editors and integrity
teams look for early signals. One signal is already in use: the ``Feet of Clay'' Detector of the Problematic Paper
Screener \citep{cabanac2022pps} spots articles whose reference lists cite retracted, withdrawn or removed papers, and
its author has posted comments on more than 1,700 such articles to prompt readers to reassess them
\citep{cabanac2024chain}. Network studies support the idea. Retracted papers cluster in citation networks
\citep{feng2024molbio,candal2024references}, and in an observational analysis of about 50,500 retracted papers in the Web of
Science, 1.43\% of the 775,000 articles citing them were themselves retracted, 12 times the global rate, rising from
0.98\% with one retracted reference to 9.58\% with five \citep{lou2026beyond}.

To the author's knowledge, that association has not been tested in the way a screening tool is used: with information frozen at one moment and
outcomes observed afterwards. A retrospective association can also arise from journals alone. Retractions arrive in
waves within journals and special issues, and papers in the same venue cite one another, so a paper that cites a
retracted paper may simply sit in a journal that is being cleaned up. A recent machine-learning study of
retraction prediction used text and metadata but no citation links, and a random rather than temporal split
\citep{fletcher2025predicting}.

This article tests the signal prospectively on open data. All predictors are computed on the January 2024
OpenCitations graph \citep{arjomandi2024dataset} with the retractions known at that time, frozen with a hash, and then
compared with retractions issued from February 2024 to October 2026, using the Retraction Watch database
\citep{rittman2025rw}. The scores, metrics and decision rule were specified in advance. The contributions are:
\begin{itemize}
\item a prospective test, with a pre-specified design, of whether citing (or being cited by) a retracted paper predicts later
retraction, on 10.5 million biomedical papers;
\item a separation of the neighbourhood signal from journal-level clustering, using strata of journal and publication
year;
\item an assessment of its value as a screen (precision and lift) against a journal baseline and a model fitted on an
earlier window;
\item supporting measurements of community concentration, NIH-funded exposure to retracted work and citations around
retraction, all from open data, with code and run records.
\end{itemize}
\section{Related work}\label{sec:related}

\paragraph{Citation of retracted papers.} Retracted papers continue to be cited after retraction, mostly without
acknowledging it: only 5.4\% of post-retraction citation contexts in PubMed Central articles mention the retraction
\citep{hsiao2021continued}, and 96\% of the analysed post-retraction citations of one fraudulent trial report did not
\citep{schneider2020continued}; see also \citet{teixeira2025citation}. Retracted papers draw more online attention than
comparable papers, and retraction does little to reduce it \citep{peng2022attention}; highly cited papers, however,
lose more citations after retraction \citep{memon2025widely}. Large open resources now document post-retraction
citations in biomedicine, including citation contexts \citep{hsiao2021continued,joseph2026sanecite}. Case studies
traced how far a single retracted result spreads through direct and indirect citations
\citep{vandervet2016propagation}, and harm has been quantified by citation distance, with indirectly citing papers
affected more than direct ones \citep{harm2025dynamics}.

\paragraph{Retractions in citation networks.} In molecular biology, the citation network of retracted papers has clear
communities, several with very high self-citation, and retracted papers and the papers citing them tend to share
retraction reasons \citep{feng2024molbio}. Of 416 paper-mill papers retracted in 2022, about 42\% cited, or were cited
by, at least one other retracted paper, mostly also from paper mills \citep{candal2024references}. Across Web of
Science, retractions cluster in citation networks, and citing retracted works is associated with higher subsequent
retraction rates \citep{lou2026beyond}. These are observational analyses of past data; none of them tests the signal
prospectively. \citet{lou2026beyond} count a citing article as retracted whenever it was retracted, and compare
with the global retraction rate rather than with papers of the same journals and years; they note that they cannot
tell whether citing retracted work raises the risk or both reflect shared factors such as paper mills.

\paragraph{Predicting retraction.} Classifiers trained on titles, abstracts and metadata reach accuracies around 0.6--0.7
on balanced samples, with a random split \citep{fletcher2025predicting}. The Feet of Clay Detector flags papers that
cite retracted work for reassessment \citep{cabanac2024chain}; to the author's knowledge, how often flagged papers are later
retracted has not been reported.

\paragraph{Funding.} Papers retracted for misconduct accounted for about \$58 million of direct NIH funding
between 1992 and 2012 \citep{stern2014costs}. After a retraction, related articles lose 5--10\% of their citations and
new articles and funding flow less into the field \citep{azoulay2015retractions}. Recent studies describe the
retractions of NIH- and NSF-funded papers themselves \citep{peterson2025nih,salisbury2026nih}. How often NIH-funded
papers cite work that has already been retracted has not, to the author's knowledge, been measured.

\paragraph{Open citation data.} OpenCitations provides open citation links and metadata
\citep{peroni2020opencitations,heibi2019coci,massari2024meta}; its coverage differs from commercial indexes
\citep{martinmartin2021coverage}.
\section{Data}\label{sec:data}

\paragraph{Citation graph.} Citations come from the parsed OpenCitations snapshot of January 2024 released through
the Illinois Data Bank \citep{arjomandi2024dataset}, built from the OpenCitations Index \citep{peroni2020opencitations,heibi2019coci}.
It holds 1,962,840,983 citation links between 107,030,533 bibliographic resources. Identifiers (DOI, PMID) and journals
were taken from the OpenCitations Meta dump \citep{massari2024meta,ocmeta2024dump}: 26,582,951 graph nodes carry a PMID
and 91,379,362 a DOI. Counting records and citing links by publication month shows that the snapshot is essentially
complete up to September 2023: 240,000--300,000 records and 8.7--10.3 million citing links per month from January to
September 2023, then 58,434 records and 2.4 million links in October and fewer afterwards.

\paragraph{Retractions.} Retraction labels come from the Retraction Watch database distributed by Crossref
\citep{rittman2025rw}. One copy was frozen for the whole study (repository commit of 30 October 2026, 73,519 records).
Only records whose nature is \emph{Retraction} are used; expressions of concern and corrections are ignored. Records
were grouped into papers by DOI, or by PMID when no DOI is given, keeping the earliest retraction notice. This gives
62,735 retracted papers. A paper is labelled \emph{paper mill} when any of its Retraction Watch reasons contains ``Paper
Mill'' (11,716 papers).

\paragraph{PubMed.} The 2026 PubMed baseline (39,994,988 records) supplies PMID--DOI links, the PubMed retraction flags
and grant agencies. A paper is \emph{NIH-funded} when one of its grants lists an agency containing ``NIH''
(3,341,220 records). A graph node is \emph{biomedical} when it has a PMID in OpenCitations Meta or its DOI appears in
PubMed (31,464,435 nodes).

\paragraph{Matching.} Retraction Watch papers were matched to graph nodes by DOI and by PMID. Of the 62,735 retracted
papers, 52,254 (83.3\%) match at least one node. Among Retraction Watch papers with a PMID, 88.8\% are also flagged as
retracted in PubMed; Retraction Watch is used as the single label source and PubMed as a check. In a random sample of 300 DOI-matched papers,
the Retraction Watch and OpenCitations titles agreed for all 292 papers that had a title in OpenCitations Meta
(differences were retraction prefixes such as ``Notice of Retraction'' or truncation).

\section{Methods}\label{sec:methods}

\subsection{Design}
The test is prospective in the sense that matters for a screening tool: all predictors use only information available
at a fixed time point $T_0$, and outcomes are retractions issued after $T_0$. The main time point is
$T_0 = $ 31 January 2024. Only 116,123 citing links in the graph (0.006\%) carry a publication date after January 2024,
so post-$T_0$ information can enter the predictors only negligibly. Because the graph is complete only to about September 2023, every analysis was
repeated with $T_0' = $ 30 September 2023. The design, scores, metrics and decision rule were written into a
specification file whose SHA-256 hash was recorded in the lab notebook before any score was computed, and the analysis
code checks the hashes of the frozen score files before reading any outcome. The specification was not deposited with a
third-party registry before the outcomes were read, so the design is pre-specified rather than formally
pre-registered; analyses added later are labelled post hoc.

\paragraph{Candidates and outcome.} Candidates are biomedical nodes published 2015--2023 that have a journal, at least
one reference in the graph, and no retraction notice dated on or before $T_0$ (10,495,850 papers). A candidate is
\emph{positive} if its first retraction notice falls between $T_0$ and 30 October 2026. Let $R_0$ denote the 44,401
graph nodes retracted on or before $T_0$. $R_0$ is defined by the date of the retraction notice; the Retraction
Watch export does not record when a notice entered the database, so a few notices dated before $T_0$ may have been
recorded only later and would not have been visible to a screen run on that date.

\paragraph{Scores.} For a candidate $p$ with $d(p)$ references in the graph:
$S_1(p)$ is the number of references of $p$ that are in $R_0$;
$S_2(p) = S_1(p)/d(p)$;
$S_3(p)$ is the number of citations $p$ received from papers in $R_0$; and
$S_4(p)$ is the number of references of $p$ to paper-mill retractions in $R_0$.
The journal baseline is a smoothed retraction rate,
\[
B(p) = \frac{r_j + m\,g}{n_j + m},
\]
where $j$ is the journal of $p$, $n_j$ is the number of biomedical items in $j$ published from 2010 to 2023,
$r_j$ is the number of those in $R_0$, $g$ is the overall rate $\sum_j r_j / \sum_j n_j$, and $m = 100$.

\subsection{Analyses}
\paragraph{Association.} Positive rates with Wilson intervals \citep{wilson1927} are reported by $S_1$ level, and risk
ratios for $S_1 \ge 1$, $S_3 \ge 1$ and $S_4 \ge 1$.

\paragraph{Within journal (primary test).} Retractions cluster by journal and publication year, for example in
mass retractions of special issues. To separate the neighbourhood signal from these clusters, the Mantel--Haenszel odds
ratio \citep{mantel1959} of exposure ($S_1 \ge 1$) on outcome was computed with strata defined by journal $\times$
publication year, with the Robins--Breslow--Greenland confidence interval \citep{robins1986estimator}. The
pre-specified decision rule was that the signal holds within journal if the 95\% interval lies above 1.

\paragraph{Screening value.} Candidates were ranked by (i) $S_1$, with ties broken by $S_2$ and then at random;
(ii) the journal baseline $B$; and (iii) a logistic model with $\operatorname{logit} B$, $\log(1+S_1)$,
$\log(1+S_3)$, an indicator of $S_1 \ge 1$ and indicators of age at the time point. The model was fitted only on an
earlier window (predictors at 31 December 2021; outcomes up to $T_0$; candidates published 2013--2021) and applied
unchanged. Precision at $k$, lift (precision divided by the base rate), recall and average precision
\citep{saito2015prc} are reported with 95\% intervals from 1,000 Poisson bootstrap resamples of the candidates.

\paragraph{Sensitivity analyses.} The analyses were repeated (a) for candidates published 2015--2022, (b) without
nodes whose DOI maps to more than one graph node, and (c) with $T_0'$.

\paragraph{Supporting analyses.} Three descriptive analyses complement the test. (1) \emph{Community concentration}:
each retracted biomedical paper with a journal was matched to one non-retracted paper from the same journal and
publication year, chosen by a fixed hash of its identifier; communities were found with the Leiden algorithm \citep{traag2019leiden} (modularity, two iterations, seeds 1
and 2) on the subgraph formed by
these papers and their neighbours, and the concentration of retracted papers over communities was compared with that
of the controls under a permutation null that swaps labels within pairs. (2) \emph{NIH exposure}: the share of
biomedical papers published 2000--2023 that cite a paper in a year after its retraction, for NIH-funded and other
papers, with a Mantel--Haenszel odds ratio stratified by journal and year. (3) \emph{Citations around retraction}:
citations per paper by year relative to the retraction year, for papers retracted 2000--2021 and matched controls.

\paragraph{Implementation.} All joins and aggregations run in DuckDB \citep{raasveldt2019duckdb} on one 8-core, 64 GB
cloud machine. Every run writes its configuration, the hash of its script, its counts and checksums; per-paper scores
are never released. The retracted papers, the citation links that touch them and the community memberships are
released as a dataset \citep{arjomandi2026retracteddata}.
\subsection{Use of generative AI}\label{sec:ai-use}
Claude Opus 5.5 (Anthropic) was used as a coding and writing assistant: to write and run analysis scripts under the
author's direction, to search the literature, to draft and edit the text, and to review drafts. All analyses, numbers
and references were checked by the author against the run records and the original sources (DOIs), and the author
takes full responsibility for the content.
\section{Results}\label{sec:results}

\subsection{Exposure and outcomes}
Of the 10,495,850 candidates, 218,561 (2.1\%) cite at least one paper that had been retracted by $T_0$, and 172,149
are cited by one. Between February 2024 and October 2026, 4,323 candidates were retracted (0.041\%), 418 of them for
paper-mill reasons.

\subsection{Association and dose}
\begin{figure}[t]
\centering
\includegraphics[width=0.6\linewidth]{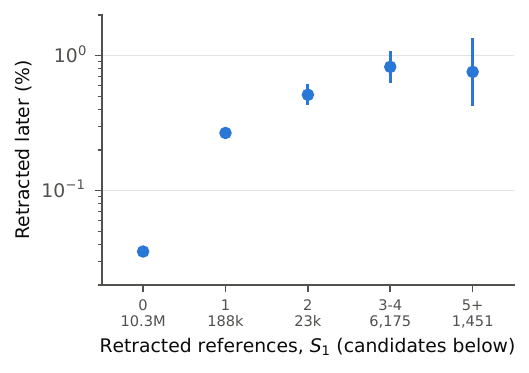}
\caption{Share of candidates retracted between February 2024 and October 2026, by the number of their references to
papers retracted by 31 January 2024 (95\% Wilson intervals; log scale). Below each level: number of candidates.}
\label{fig:dose}
\end{figure}

Candidates that cite a retracted paper were retracted later far more often (Table~\ref{tab:dose}, Fig.~\ref{fig:dose}). The rate rises from
0.035\% without a retracted reference to 0.27\% with one, 0.51\% with two and 0.83\% with three or four; the
1,451 candidates with five or more retracted references show 0.76\%, with a wide interval. The risk ratio for
$S_1 \ge 1$ is 8.8 (95\% CI 8.1--9.6). It is 17.8 (15.5--20.6) for citing a paper-mill retraction, and 6.8
(6.2--7.5) for being cited by a retracted paper.

\begin{table}[t]
\centering
\caption{Later retraction by number of references to papers retracted by $T_0$ (main analysis).}\label{tab:dose}
\begin{tabular}{lrrr}
\toprule
$S_1$ & Candidates & Retracted later & Rate (95\% CI) \\
\midrule
0 & 10,277,289 & 3,640 & 0.035\% (0.034--0.037) \\
1 & 187,735 & 502 & 0.27\% (0.25--0.29) \\
2 & 23,200 & 119 & 0.51\% (0.43--0.61) \\
3--4 & 6,175 & 51 & 0.83\% (0.63--1.08) \\
$\ge 5$ & 1,451 & 11 & 0.76\% (0.42--1.35) \\
\bottomrule
\end{tabular}
\end{table}

\subsection{Within journal and year}
\begin{figure}[t]
\centering
\includegraphics[width=0.95\linewidth]{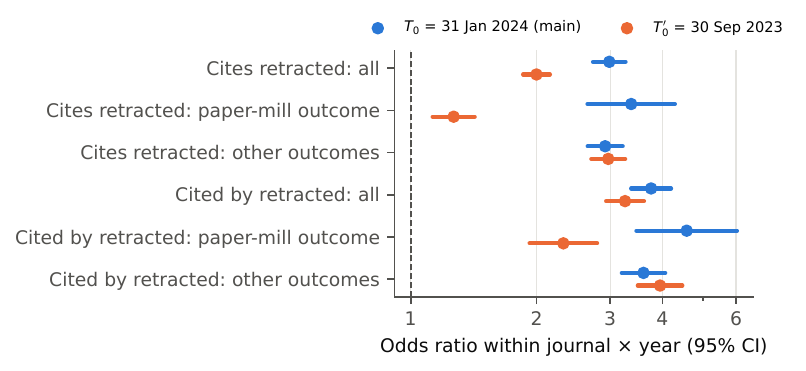}
\caption{Mantel--Haenszel odds ratios within journal $\times$ publication year (95\% Robins--Breslow--Greenland
intervals), for citing a retracted paper ($S_1 \ge 1$) and being cited by one ($S_3 \ge 1$), with the main and the
earlier time point.}
\label{fig:mh}
\end{figure}

Part of this association reflects journals: retractions cluster in journals and publication years, and papers in the
same journal cite one another. Within journal $\times$ publication year, the Mantel--Haenszel odds ratio for
$S_1 \ge 1$ is 2.99 (95\% CI 2.73--3.27; 1,201 informative strata holding 4,258 of the positives). By the
pre-specified rule, the signal holds within journal (Fig.~\ref{fig:mh}). It holds for paper-mill outcomes (3.37, 2.65--4.29) and for
other outcomes (2.92, 2.65--3.21). Being cited by a retracted paper gives a within-journal odds ratio of 3.76
(3.37--4.20). The estimate is unchanged without candidates whose DOI maps to several graph nodes (2.99) and slightly
higher for papers published 2015--2022 (3.10, 2.82--3.41).

\subsection{Reference-list length and reviews (post hoc)}
Papers that cite a retracted paper have longer reference lists (median 51 vs 34 references), so longer lists alone
could raise both the chance of citing a retracted paper and of being retracted. Adding deciles of the number of
references to the journal $\times$ year strata leaves the odds ratio essentially unchanged: 3.03 (2.76--3.33; 1,838
strata), 4.12 (3.21--5.29) for paper-mill outcomes and 2.89 (2.61--3.19) for others. The share of retracted references
($S_2$) shows the same gradient: within journal, year and reference-length decile, the odds ratio against papers
with no retracted reference is 1.74 (1.14--2.66) for a share up to 1\%, 2.82 (2.33--3.41) for 1--2\%, 2.98
(2.64--3.35) for 2--5\% and 3.96 (3.24--4.85) above 5\%.

Reviews explain the flat rate at the top of Table~\ref{tab:dose}. Using PubMed publication types (review, systematic
review, meta-analysis), reviews make up 12\% of candidates without a retracted reference but 48\% of those with three
or four and 70\% of those with five or more, and they are almost never retracted (0.1\% at five or more). Among the
other papers the rate rises steadily: 0.036\% with no retracted reference, 0.35\% with one, 0.79\% with two,
1.52\% with three or four and 2.27\% (10 of 441; 95\% Wilson interval 1.24--4.12) with five or more. Without reviews, the within-journal odds ratio is
3.19 (2.90--3.51), and the screen improves: among the 1,000 non-review papers with the most retracted references, 22
were retracted later (2.2\%, 95\% Wilson interval 1.5--3.3; lift 52), against 9 (0.9\%) when reviews are kept in
the ranking. These analyses were added after the outcomes had been seen and are labelled post hoc.

\subsection{Screening value}
Despite the strong relative effect, the signal is weak as a screen (Table~\ref{tab:screen}). Most later retractions,
3,640 of 4,323 (84\%), cite no retracted paper at all. Among the 1,000 candidates with the most retracted references,
9 were retracted later (0.9\%, lift 22 over the base rate); among the top 10,000, 72 (0.72\%), which is 1.7\% of all
positives. The journal rate alone ranks one positive among its top 1,000 (0.1\%). The combined model, fitted on the earlier
window, has the highest average precision (0.0037) but a lower precision at 1,000 than the $S_1$ ranking.

\begin{table}[t]
\centering
\caption{Screening performance in the main analysis (95\% bootstrap intervals). Base rate 0.041\%.}\label{tab:screen}
\begin{tabular}{lrrrr}
\toprule
Ranking & Prec.@1,000 & Lift@1,000 & Prec.@10,000 & Avg. precision \\
\midrule
$S_1$ & 0.90\% (0.40--1.50) & 21.9 & 0.72\% (0.56--0.89) & 0.0013 \\
Journal rate $B$ & 0.10\% (0--0.30) & 2.4 & 0.02\% & 0.0023 \\
Combined model & 0.40\% (0.10--0.80) & 9.7 & 0.79\% (0.61--0.94) & 0.0037 \\
\bottomrule
\end{tabular}
\end{table}

\subsection{When the time point falls inside a retraction wave}
With $T_0' = $ 30 September 2023, the outcome window contains the mass retractions of late 2023 (7,578 positives,
3,141 of them paper-mill). The journal rate then becomes the best screen: 6.8\% of its top 1,000 candidates were
retracted (lift 94), against 1.1\% for $S_1$. The within-journal odds ratio for $S_1 \ge 1$ falls to 2.00
(1.86--2.15), and to 1.27 (1.13--1.42) for paper-mill outcomes, while it stays at 2.97 for other outcomes. By
January 2024 those waves had passed, and the journal rate placed only one future retraction among its top 1,000.

\subsection{Supporting analyses}
\paragraph{Community concentration.} The 27,567 retracted biomedical papers with a journal, their controls from the
same journal and year, and all their neighbours form a subgraph of 2.7 million papers and 57 million citations. Leiden
found 197 and 200 communities with two seeds (modularity 0.65 and 0.66); the two partitions agree only moderately
(adjusted Rand index 0.41), so both are reported. Retracted papers are more concentrated than their controls: the probability that
two retracted papers share a community is 1.15 and 1.11 times that of two controls (permutation $p = 0.001$ for
both, with a null mean of 1.00). The excess comes mainly from paper-mill retractions (1.68 and 1.52); for other
retractions it is small (1.06 and 1.05). Retracted paper-mill papers thus cluster in the citation graph beyond what
their journal and year imply, in line with \citet{feng2024molbio} and \citet{candal2024references}.

\paragraph{NIH exposure.} Of 2,332,727 NIH-funded biomedical papers published 2000--2023 in the graph, 10,872 (0.47\%)
cite at least one paper in a year after its retraction, compared with 0.48\% of other biomedical papers. Within journal
and year, NIH funding makes no practical difference (odds ratio 1.03, 1.00--1.05). NIH-funded papers cited 3,548
distinct retracted papers after their retraction.

\begin{figure}[t]
\centering
\includegraphics[width=0.6\linewidth]{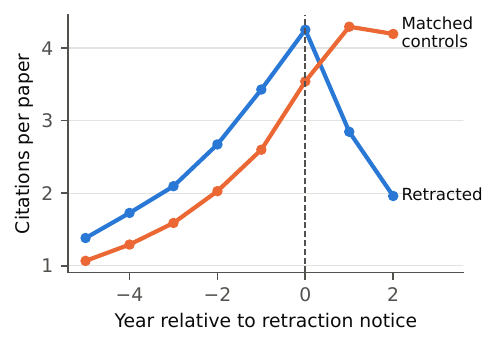}
\caption{Citations per paper by year relative to the retraction notice, for 11,788 papers retracted 2000--2021 and
controls matched on journal and publication year.}
\label{fig:event}
\end{figure}

\paragraph{Citations around retraction.} For 11,788 papers retracted 2000--2021 and matched controls from the same
journal and year, retracted papers received about 1.3 times the controls' citations in each of the five years before
the retraction year, 0.66 times in the year after and 0.47 times two years after (Fig.~\ref{fig:event}). Citations continue after
retraction, at about two per paper in the second year, consistent with earlier studies
\citep{hsiao2021continued,schneider2020continued}.

\section{Discussion}\label{sec:discussion}

\paragraph{The signal is real and not just the journal.} Citing a paper that has already been retracted is followed by a
markedly higher chance of retraction in the next three years, and the association survives comparison within journal
and publication year (odds ratio about 3). This turns the retrospective association reported by
\citet{lou2026beyond}, who compared citing articles with the global retraction rate, into a prospective one. Part of
the crude association does reflect journals: the crude risk ratio of 8.8 shrinks to an odds ratio of 3.0 within journal
and year (for an outcome this rare the two measures are nearly equal). But a threefold excess remains, so the signal is not explained by retraction waves within journals, and holding
reference-list length fixed does not change it. Their rates are also higher than ours (9.58\% at five retracted
references against 2.27\% here for non-review papers): they count retractions over the whole history of each citing
article and include articles retracted together with the papers they cite, whereas here only retractions within 2.75
years after $T_0$ count and every paper already retracted by $T_0$ is excluded.
The rate rises with the number of retracted references, and among non-review papers it keeps rising up to five or
more; citing a paper-mill retraction is the strongest single marker. Being cited by a retracted paper is at least as informative as citing one, which fits the
picture of interlinked clusters of problematic papers \citep{candal2024references}.

\paragraph{But it is a weak screen on its own.} A relative risk of nine on a base rate of 0.04\% still means that almost
all flagged papers are not retracted: about 1 in 110 of the top 1,000 by $S_1$. And 84\% of later retractions cite no
retracted paper at all, so the signal misses most cases. A ``Feet of Clay'' flag is therefore a reason to look, not
evidence of a problem. Many papers cite retracted work for good reasons, such as reviews, critiques, or citations made
before the retraction. Reviews are a clear case: they make up most of the papers with many retracted references and are almost
never retracted, so a screen should set them aside, which raised precision among the top 1,000 from 0.9\% to 2.2\%; among other papers the risk keeps rising with the number of
retracted references.

\paragraph{Timing matters.} The best screen depends on where the time point falls relative to retraction waves. Inside
the late-2023 wave, the journal's own retraction rate identified future retractions with a lift of about 90, and the
neighbourhood signal added little for paper-mill cases. Once the wave had passed, the journal rate carried almost no
information about who would be retracted next, while the neighbourhood signal kept its strength. A practical screen
should combine both and be refitted as waves come and go; the model fitted on outcomes up to January 2024, which put weight on the journal
rate, ranked fewer later retractions at the very top than the reference count alone.

\paragraph{NIH-funded research.} NIH-funded papers cite already-retracted work as often as other papers in the same
journal and year. About 0.5\% of biomedical papers published 2000--2023 do so: 103,349 papers in the graph, 10,872 of them
NIH-funded. Whether these citations rely on the retracted findings cannot be read from
citation links alone; citation-context resources \citep{joseph2026sanecite,hsiao2021continued} could answer it.

\paragraph{Limitations.} The test is limited to papers in the January 2024 OpenCitations snapshot, which depends on
publishers depositing references with Crossref and is complete only to about September 2023; papers without deposited
references cannot be scored, and 17\% of retracted papers could not be matched to the graph. Retraction Watch is
incomplete: 3,205 papers flagged as retracted in PubMed are missing from the copy used, three quarters of them with
notices from 2024 onwards, so some later retractions are missing from the outcome labels. A retraction is not proof of
misconduct; non-retracted candidates include papers that should have been
retracted, so precision is a lower bound on the share of problematic papers among the flagged ones. The outcome window
is under three years, while the median time from publication to retraction for biomedical papers retracted after January
2024 was about 2.7 years, so
many later retractions are not yet observed. NIH funding is taken from PubMed grant lists, which are incomplete. All
results are associations.
\section{Conclusion}\label{sec:conclusion}
On 10.5 million biomedical papers, citing a paper that had already been retracted predicted later retraction
prospectively and within journal and year, with an odds ratio of about 3. The signal is a useful flag but a weak
screen on its own: it ranks a few later retractions highly and misses most of them. Integrity screening should treat
links to retracted work as one input, alongside journal-level signals that matter most during retraction waves. All
data are open and the analysis can be rerun from the released code and records.

\backmatter

\section*{Declarations}\label{sec:availability}

\begin{itemize}
\item \textbf{Funding:} No funding was received for this work.
\item \textbf{Competing interests:} The author has no competing interests to declare.
\item \textbf{Ethics approval and consent to participate:} Not applicable. The study uses only public bibliographic
records; no author names are analysed or released.
\item \textbf{Consent for publication:} Not applicable.
\item \textbf{Data availability:} %
Citation data: \citet{arjomandi2024dataset}; identifiers and journals: \citet{ocmeta2024dump}; retractions: the
Retraction Watch database distributed by Crossref (repository commit of 30 October 2026); PubMed 2026 baseline courtesy
of the U.S. National Library of Medicine (it may not reflect the latest NLM data). Scripts, configuration, the
specification file, run records with checksums, and the aggregate result tables:
\url{https://github.com/HosseinMA96/retracted-neighbours-open-data}. The repository includes the specification file
with its recorded hash, a step-by-step reproduction guide, and a map from every number in this article to the run that
produced it. The derived dataset (all 62,735 retracted papers with retraction dates and reasons, the 2.26 million citation
links that touch them, and the community memberships; no per-paper scores for non-retracted papers, no author names) is
on Zenodo \citep{arjomandi2026retracteddata}: \url{https://doi.org/10.5281/zenodo.23874477}.
\item \textbf{Code availability:} See Data availability.
\item \textbf{Author contributions:} Hossein Mohasel Arjomandi is the sole author and designed the study, ran the analyses and wrote the manuscript.
\end{itemize}

\bibliography{refs}

\end{document}